\documentclass[published]{JHEP3} 

\usepackage{epsfig,multicol}

\newcommand{\half}{{1\over2}}

\newcommand{\ie}{{\em i.e.}\ }
\newcommand{\eg}{{\em e.g.}\ }
\newcommand{\efolds}{$e$-folds\ }
\newcommand{\Mp}{M_{\rm P}}
\newbox\pippobox

\title{Inflationary scenarios from branes at angles}

\author{Juan Garc\'{\i}a-Bellido, Ra\'ul Rabad\'an~and Frederic Zamora\\
        Theory Division CERN, CH-1211 Gen\`eve 23, Switzerland\\
        E-mail: \email{bellido@mail.cern.ch}, \email{Raul.Rabadan@cern.ch}, 
     \email{Frederic.Zamora@cern.ch}}
\received{January 17, 2002}               
\accepted{January 24, 2002}               

\JHEP{01(2002)036}

\preprint{\hepth{0112147}}      
\preprint{CERN-TH/2001-372}      
\preprint{IFT-UAM/CSIC-01-40}

\abstract{
We describe a simple mechanism that can lead to inflation within
string-based brane-world scenarios. The idea is to start from a
supersymmetric configuration with two parallel static Dp-branes, and
slightly break the supersymmetry conditions to produce a very flat
potential for the field that parametrises the distance between the
branes, \ie the inflaton field. This breaking can be achieved in
various ways: by slight relative rotations of the branes with small
angles, by considering small relative velocities between the branes,
etc. If the breaking parameter is sufficiently small, a large number
of \efolds can be produced within the D-brane, for small changes of
the configuration in the compactified directions. Such a process is
local, \ie it does not depend very strongly on the compactification
space nor on the initial conditions. Moreover, the breaking induces a
very small velocity and acceleration, which ensures very small
slow-roll parameters and thus an almost scale invariant spectrum of
metric fluctuations, responsible for the observed temperature
anisotropies in the microwave background. Inflation ends as in hybrid
inflation, triggered by the negative curvature of the string tachyon
potential. In this paper we elaborate on one of the simplest examples:
two almost parallel D4-branes in a flat compactified space.
}


\begin{document}

\section{Introduction}

Inflation is a paradigm in search of a model~\cite{Lindebook}. It has
been for several years the aim of particle physicists to construct
models of inflation based on supersymmetry (in search of sufficiently
flat potentials) and string theory (in hope of a description within
quantum gravity)~\cite{LythRiotto}.  It is thus worthwhile the
exploration of the various possibilities present within string theory,
to come up with cosmological models consistent with observations.

In this respect, intersecting brane systems have very interesting
features; see, among others, Refs.~\cite{bdl96,general}.  For instance,
they provide constructions that are very close to the standard model,
even with the same spectrum of particles
\cite{bgkl00,bkl00,afiru001,imr01,csu01,bklo01}. In general these are
non-supersymmetric, but there are also some supersymmetric
constructions~\cite{csu01}. Within string theory, the interaction
between the branes arises due to the exchange of massless closed string
modes. When supersymmetry is preserved, the Neveu-Schwarz-Neveu-Schwarz
(NS-NS) and Ramond-Ramond (R-R) charges cancel and there is no net force
between the branes.

Recently, the proposals of
Refs.~\cite{dt98,dss01,bmnqrz01,hhk01,klps01,ks01,h01,st01} have
derived some of the inflationary properties from concrete
(non-supersymmetric) brane configurations. In this paper we will show
that inflation is a very general feature for non-supersymmetric
configurations which are not far away from the supersymmetric
one.\footnote{The same happens in Refs.~\cite{hhk01,ks01,h01}.}  The
idea is to break slightly the supersymmetric configuration so one can
smoothly turn on an interaction between the branes. Inflation occurs
as the branes are attracted to each other, but a tachyonic instability
develops when the branes are at short distances compared with the
string scale. To our knowledge, this property of ending inflation like
in hybrid models through the open string tachyon was first proposed in
Ref.~\cite{bmnqrz01}. This is a signal that a more stable
configuration with the same R-R total charge is available, which
by itself triggers the end of inflation.

There are many ways in which this general idea can be implemented: by
a slight rotation of the branes intersecting at small angles
(equivalent to adding some magnetic fluxes in the T-dual picture), by
considering small relative velocities between the branes, etc. In this
paper we elaborate on one of the simplest examples: a pair of
D4-branes intersecting at a small angle in some compact directions.
The interaction can be made arbitrarily weak by choosing the
appropriate angle to be sufficiently small. The brane-antibrane system
is the extreme supersymmetry-breaking case, where the angle is
maximised such that the orientation for one brane is opposite to the
other. The interaction is so strong in this case, that inflation seems
hard to realise. One should take very particular initial conditions on
the system for inflation to proceed. On the other hand, if the
supersymmetry breaking parameter is small, a huge number of \efolds
are available within a small change of the internal configuration of
the system, due to the almost flatness of the potential; in this way
the initial conditions do not play an important role.  Thus, we find
that inflation appears very naturally in systems that are not so far
from supersymmetry preserving ones.

In this paper we will mainly focus on one of the simplest realisations
of intersecting branes and extract the first consequences for inflation.
In section 2 we will describe the system and its decay, due to the
tachyon instability, to a supersymmetric brane.  We devote section 3 to
derive the effective action for the inflaton at large distances with
respect to the string length scale; we present two different (although
equivalent) ways to obtain the action.  In section 4 we explain how
inflation is produced in this model, indicating explicitly the
conditions that a generic model should satisfy for producing a
successful cosmological scenario. After that, in section 5, we discuss
the behaviour of the system for inter-brane distances of the order of
the string scale by using the description of the low energy effective
field theory living on the branes. In section 6 we discuss very briefly
the conditions that give rise to an efficient reheating within our world
brane and derive the reheating temperature.  The last section is devoted
to the conclusions.

The paper comes with three complementary Appendices, where concrete
computations are done for the inflaton's effective action (Appendices
A and B) and the transverse space compactification effects (Appendix
C).

\section{Description of the model}

Consider type IIA string theory on ${\cal R}^{3,1}\times T^6$, with
$T^6$ a (squared) six torus.  Let us put two D4-branes expanding 3+1
world-volume dimensions in ${\cal R}^{3,1}$, with their fourth spatial
dimension wrapping some given 1-cycles of $T^6$. In
Fig.~\ref{1angleD4} we have drawn a concrete configuration.

\EPSFIGURE[ht]{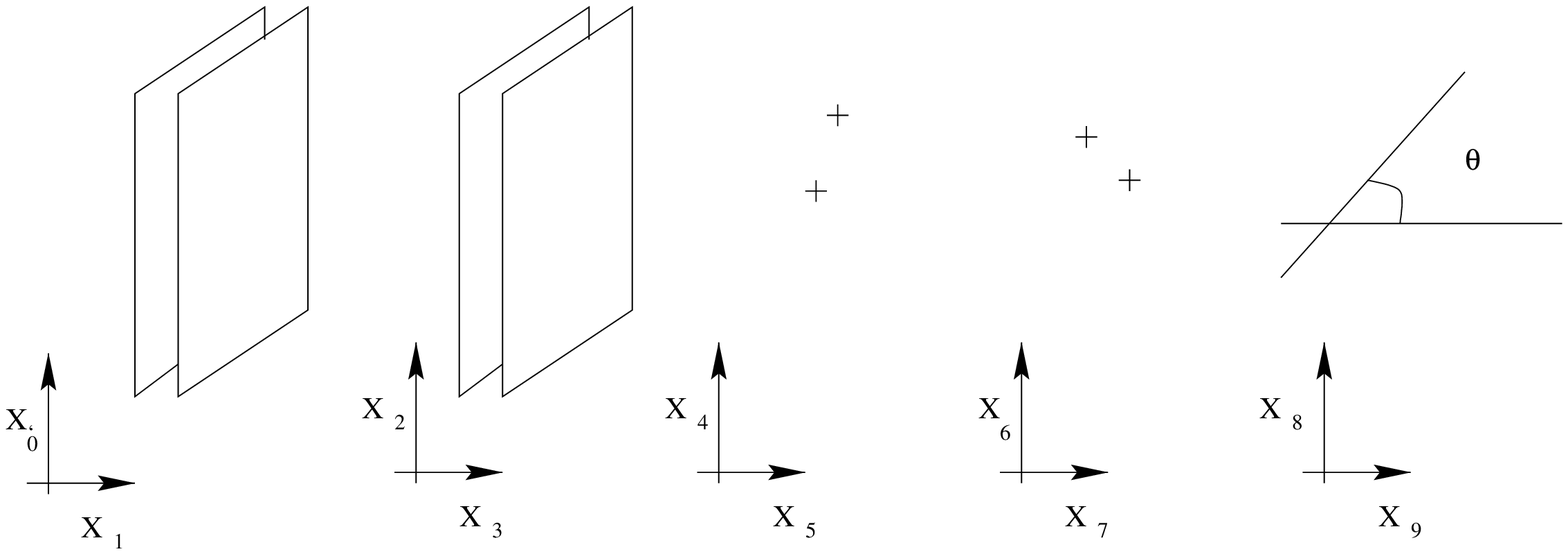,width=12cm}{\label{1angleD4} This figure
represents two D4-branes at an angle $\theta$.  The usual
4-dimensional spacetime is along directions $\{x^0,...,x^3\}$. The
branes are located at particular points on the compact four directions
$\{x^4,...,x^7\}$. Finally, they wrap different cycles on the last two
compact directions, $x^8$ and $x^9$, and intersect at a given angle
$\theta$.}

If both branes are wrapped on the same cycle and with the same
orientation, we have a completely parallel configuration that preserves
sixteen supercharges, \ie ${\cal N}=4$ from the 4-dimensional point
of view.  If they wrap the same cycle but with opposite orientations, we
have a brane-antibrane configuration, where supersymmetry is completely
broken at the string scale \cite{bs95}.  But for a generic configuration
with topologically different cycles, there is a non-zero relative angle,
let us call it $\theta$, in the range $0\leq \theta \leq \pi$, and the
squared supersymmetry-breaking mass scale becomes proportional to
$2\theta/(2\pi \alpha')$, for small angles.  The case we are considering
can be understood as an intermediate case between the supersymmetric
parallel branes ($\theta =0$) and the extreme brane-antibrane pair
($\theta =\pi$), with the angle playing the role of the smooth
supersymmetry breaking parameter in units of the string length.

Notice that this configuration does not satisfy the R-R tadpole
conditions \cite{pc88}. These conditions state that the sum of the
homology cycles where the D-branes wrap must add up to zero. In our
case these conditions do not play an important role, since we can take
a brane, with the opposite total charge, far away in the transverse
directions. This brane will act as an {\em expectator} during the
dynamical evolution of the other two branes.  We will come back to
this issue of R-R tadpole cancellation with an expectator brane when
discussing reheating.  Also, since the configuration is
non-supersymmetric, there are uncancelled NS-NS tadpoles that should
be taken into account \cite{fs86,dm00,bf00,p98}. They act as a
potential for the internal metric of the manifold, \eg the complex
structure of the last two torus in the 8-9 plane,\footnote{See for
  example \cite{bklo01}, where NS-NS tadpoles are analysed in the
  context of intersecting brane models.} and for the dilaton. Along
this paper we will consider that the evolution of these closed string
modes is much slower than the evolution of the open string modes.

\EPSFIGURE[ht]{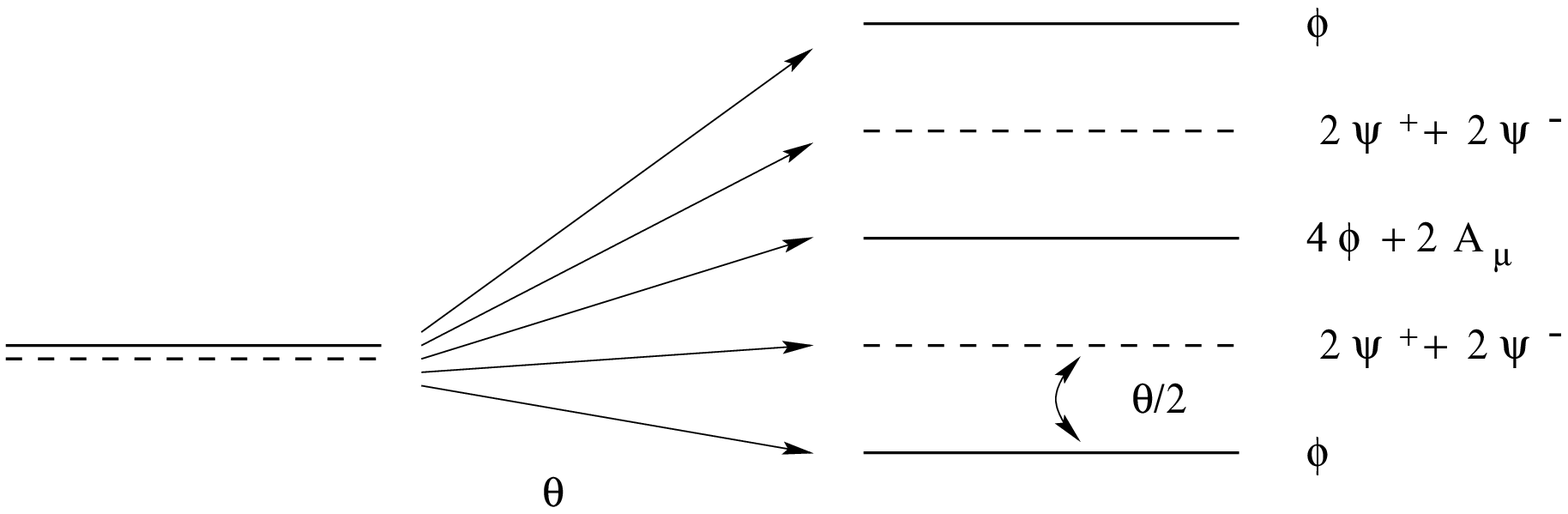,width=10cm}{\label{split} Splitting of the mass
spectrum of the ${\cal N} = 4$ Yang-Mills supermultiplet when the angles
are non-vanishing or, in the T-dual picture, degenerancy splitting for
the first Landau level.  $\psi^{\pm}$ are the chiral fermions and $\phi$
are the scalars.}

Each brane is located at a given point in the two planes determined by
the compact directions $\{x^4,...,x^7\}$; let us call these points
$y_i$, $i=1,2$. In the supersymmetric ($\theta=0$) configuration there
is no force between the branes and they remain at rest. When $\theta
\not=0$, a non-zero interacting force develops, attracting the branes
towards each other. From the open string point of view, this force is
due to a one-loop exchange of open strings between the two branes.
The coordinate distance between the two branes in the compact space,
$y^2 = |y_1 - y_2|^2$, plays the role of the inflaton field, whose
vacuum energy drives inflation.

Table~\ref{mass1} gives the spectrum for the lightest open string
states between the branes. The open string spectrum can be obtained
from the quantization of open strings with ends attached to the
D-branes at an angle, or in the T-dual picture by considering a
magnetic flux on the brane. Besides the usual contribution to the mass
from the $y^2 \not=0$ expectation value, there is a non-trivial
splitting for the whole ${\cal N}=4$ supermultiplet, proportional to
the angle $\theta$. It is interesting to point out the origin of this
splitting from the point of view of the field theory living on the
brane.  We start from two D4-branes, one wrapping the cycle $(0,1)$
and the other the cycle $(1,N)$ of the squared torus in $\{x^8,x^9\}$.
If we T-dualise in the $x^8$ direction, we obtain a 6-dimensional
${\cal N}=2$, $U(N+1)$ Super Yang-Mills theory living on the D5
branes.  The distance between the branes triggers a Super-Higgs
mechanism of $U(N+1) \to U(N) \times U(1)$. The non-trivial
homological charge due to the wrapping in the $x^8$ direction gives
one unit of magnetic flux through the dual torus, for the $U(1)
\subset U(N)$ factor, $F_{N} = H {\cal I}_N$, where we have
$$
H = {1\over2\pi NR'R} = {1\over2\pi\alpha'N} = {\theta\over2\pi\alpha'} \,.
$$
The magnetic field $H$ couples to the corresponding off-diagonal
fluctuations of the $N+1$ adjoint matrices representing the charged
particles with respect to this Abelian field. One can compute their
mass spectrum on the 4-dimensional reduced field theory by finding the
eigen-states of the operator \cite{bachas}
\begin{equation}
{\cal M}^2_H = (p_8 -iA_8)^2 + (p_9 -iA_9)^2 + 2H{\cal S}_{89}\,,
\end{equation} 
with ${\cal S}_{89}$ the spin operator on the 8-9 toric plane. Since the
magnetic field is constant, the operator ${\cal M}^2$ gives a spectrum
of Landau levels, with a spin-dependent splitting for the 6-dimensional
${\cal N}=2$ vector multiplet at each level, see Fig.~\ref{split}. We
have the eigenstates
\begin{equation}
\label{mass_spectrum}
m^2_{H,n} = (2n+1)H + 2H{\cal S}_{89}\,, \hspace{1cm} n=0,1,2,...
\end{equation}
The 6-dimensional ${\cal N}=2$ vector multiplet contains four scalars,
two opposite chiral fermions and one massless vector.  A chiral fermion
in six dimensions becomes a Dirac fermion in four dimensions. From
(\ref{mass_spectrum}), its two Weyl components get splitted depending on
the sign of ${\cal S}_{89} = \pm\half$. Out of the four physical degrees
of freedom of the massless vector, only those two with the spin in the
8-9 plane, with ${\cal S}_{89} = \pm 1$, are splitted. From the
4-dimensional point of view, they are the two scalars of the ${\cal
N}=2$ vector multiplet.  With respect the remaining four scalars, which
belong to the 4D ${\cal N}=2$ hypermultiplet, one of them is mixed with
the two vectorial degrees of freedom with spin ${\cal S}_{89} =\pm 1$ to
give a 4D massive spin 1 field.  As one can check, after equating the
angle, $\theta = 1/N$, the lowest Landau level reproduces the spectrum
of the lightest open strings given in Table 1.

Apart from this low-energy supermultiplet, there are copies of these
particles at higher levels, separated by a mass gap $\Delta m^2 =
2\theta/(2\pi\alpha')$; these states were called gonions in
Ref.~\cite{afiru001}. In the T-dual picture, where an angle
corresponds to a magnetic flux perpendicular to the brane, all these
states correspond to higher Landau levels~\cite{bachas,bdl96}, see
Eq.~(\ref{mass_spectrum}). In the next section we will describe their
connection to the string states. Note that, for each Landau level, not
only the supertrace of the squared masses cancel, but also the sum of
their masses up to the sixth power:
\begin{equation}
\sum_i (-1)^{F_i}\,m_i^{2n}=0\,, \hspace{2cm} n=1,2,3\,.
\end{equation}
This is because supersymmetry is spontaneously broken 
by the non-zero magnetic flux in the T-dual 6-dimensional 
Super Yang-Mills theory.

\TABLE[h]{
\begin{tabular}{|c|c|} \hline
Field & $2\pi\alpha'm^2 - \frac{y^2}{2\pi\alpha'}$ \\
\hline
\hline
1 scalar &  $3 \theta$ \\
\hline 
2 massive fermions &   $2\theta$ \\
\hline 
3 scalars &   $\theta$ \\
\hline 
1 massive gauge field &   $\theta$ \\
\hline 
2 massive fermions (massless for $y=0$) &   $0$ \\
\hline 
1 scalar (tachyonic for $y=0$) &   $- \theta$ \\
\hline 
\end{tabular}\label{mass1}
\caption{The mass spectrum of the ${\cal N} = 4$ supermultiplet.}
}

Notice that the first scalar will be tachyonic if the distance between
the two branes is smaller than $y_c^2 = 2\pi\alpha'\theta$, \ie
the two-brane system becomes unstable.  It can minimise its volume
(and therefore its energy) by decaying to a single brane with the same
R-R charges as the other two but with lower volume, see
Fig.~\ref{decay}.  Sen's conjecture ~\cite{Sen99} relates the
difference between the energy of the initial and final states with the
change of the tachyonic potential:
\begin{equation}\label{senconj}
\Delta V_{\rm tachyon} = T_4 (V_1 + V_2 - V_f)\,,
\end{equation}
where $T_p = M_s^{p+1} g_s^{-1}/(2\pi)^p$ is the tension of the
branes, with $M_s=(\alpha')^{-1/2}$ the string mass and $g_s$ the
string coupling; $V_i$ are the brane world-volumes, which take into
account the possible wrappings around the compactified space.

\EPSFIGURE[ht]{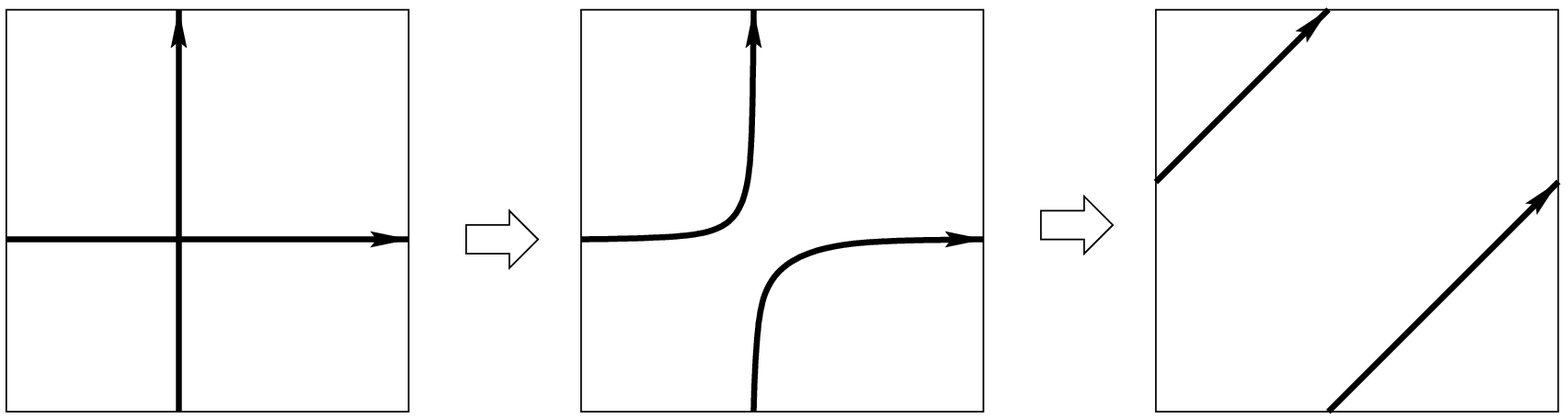,width=12cm}{\label{decay}Decay of a two brane 
system on a $T^2$.}
 
To visualise how this process happens let us take a simple toy model.
Consider two branes wrapping 1-cycles in a two dimensional squared
torus of radius R. Each of these branes wraps a straight line in the
$\{n_i [a] + m_i [b]\}$ homology class. The energy density of the
two-brane system is just \cite{p98}
$$E_0 = T_4 R \Big[\sqrt{n_1^2 + m_1^2} + \sqrt{n_2^2 +
m_2^2}\Big]\,.$$ These two branes have a total (homological) charge
$\{(n_1 + n_2) [a] + (m_1 + m_2) [b]\}$. Consider for example two cycles
with angles $\phi_1={\rm arctan} (m_1/n_1)\equiv \theta$ and
$\phi_2={\rm arctan}(m_2/n_2)=0$, so that the angle between the two
branes is $\theta=|\phi_1-\phi_2| \ll 1$. In this case we can write
the initial energy density as
\begin{equation}
E_0 \simeq 2T_4 L\,\Big(1 + {n_1\tan^2\theta \over 2(n_1+n_2)} +
{\cal O}(\theta^4)\Big)\,,
\end{equation}
where $2L=R(n_1+n_2)$ is the length of the brane wrapped around the
two cycles.  However, there is a brane configuration that has the same
charges but less energy, see Fig.~\ref{decay}. This is a brane
wrapping a straight line in the $\{(n_1 + n_2) [a] + (m_1 + m_2)
[b]\}$ homology class. In this case, the energy density of the system
is:
\begin{equation}
E_f = T_4 R \sqrt{(n_1 + n_2)^2 + (m_1 + m_2)^2}\ 
\simeq 2T_4 L\,\Big(1 + {n_1^2\tan^2\theta \over
2(n_1 + n_2)^2} + {\cal O}(\theta^4)\Big)\,.
\end{equation}
From Sen's conjecture (\ref{senconj}), we have
\begin{equation}
\Delta V_{\rm tachyon}
 = E_0 - E_f 
\simeq T_4 R \,\tan^2\theta\,{\mu\over2}\,,
\end{equation}
where $\mu=n_1n_2/(n_1+n_2)$ is the ``reduced'' winding number of the
two branes. It is interesting to note that in the case of a brane-brane
system ($\theta =0$) we obtain $\Delta V_{\rm tachyon} = 0$, as
expected, while in the case of brane-antibrane (where supersymmetry is
broken at the string scale) we have $\Delta V_{\rm tachyon} = 2T_4L$,
since $E_f=0$.  In our case, since we are breaking supersymmetry only slightly
($\theta\ll1$), the energy difference is proportional to $M_{\rm susy}^4
\sim (\theta/ 2\pi\alpha')^2$.  In the small angle approximation we have
\begin{equation}\label{V0}
\Delta V_{\rm tachyon} \simeq 2T_4 L\,{\tan^2\theta\over8} \simeq
{M_sL\,\over 16\pi^2 g_s}\,\Big({\theta\over2\pi\alpha'}\Big)^2\,.
\end{equation}
This corresponds to the energy difference between the false vacuum with
$E_0 = 2T_4L$, and the final true vacuum at the minimum of the tachyon
potential.\footnote{Note that the argument can be straightforwardly
generalised to an arbitrary number of branes. In higher dimensions, 
\ie branes wrapping higher dimensional manifolds that intersect at
some points, the stability can be analysed in a similar way
\cite{l89,d99,r01}. The absence of tachyons depends on the angles of the
system at the intersecting point. In these cases, branes can intersect
without preserving any supersymmetry and still being non-tachyonic.}

\section{Supergravity description at long distances}

When the two branes are at a distance much larger than the string
scale, \ie  $y\gg l_s$, the effective action for the inflaton
field $y$ can be computed from the exchange of massless closed string
modes.  In this section, we will present two different (although
equivalent) ways to obtain the inflaton effective action.


One can compute the closed string tree-level interaction between two
D-branes by going to the open string dual channel. In that case, the
interaction potential corresponds to the one-loop vacuum amplitude for
the open strings \cite{as96,p98},
\begin{eqnarray}\label{potD4}
V(y_i,\theta) &=& - V_4 \int_0^\infty \frac{dt}{t}(8 \pi^2 \alpha'\,t)^{-2} 
 \sum_{m_i} e^{- {t\over2\pi\alpha'} 
\sum_i (y_i + m_i 2\pi R_i)^2} Z(\theta,t)\,,\\
Z(\theta,t) &=& {\theta^4_{11}(i \theta t/2\pi,it)\over
i\theta_{11}(i \theta t/\pi,it)\eta^9(it)} \,,  \label{ZD4}
\end{eqnarray}
where $\eta(\tau)$ is the Dedekind eta function, and
$\theta_{11}(\nu,\tau)$ are the Jacobi elliptic functions~\cite{p98}.
Here $m_i$ are the winding modes of the strings on the $\{X_i,\
i=4,\dots,7\}$ transverse directions, \ie the Dirichlet-Dirichlet
ones, see Fig.~\ref{1angleD4}. The $R_i$ are the radii of these
compact dimensions and the $y_i$ are the distances between the branes
in each of these compact dimensions.  The prefactor $V_4$ is the
regularised volume of our 4 Minkowski coordinates. To get the
vacuum energy density one should take the ratio over the 4-volume,
$V(y_i)/V_4$.

Let us explain briefly how this potential arises. The first factor
comes from the integration over momenta in the non-compact
dimensions. The sum over the integer numbers $m_i$ comes from the
winding modes in the compact transverse directions.  The $\theta_{11}$
and $\eta$ elliptic functions come from the bosonic and fermionic
string oscillators of the world sheet, see Ref.~\cite{p98}.  Notice
also that the potential (\ref{potD4}) is not invariant under rotations
in the compact coordinates. This is due to the toroidal
compactification. We will see that in the limit where the
compactification scale is much greater than the distance $y$ between
branes, the potential becomes invariant under the group of rotations
$SO(4)$ in these coordinates.

\begin{figure}[ht]\begin{center}
\epsfig{file=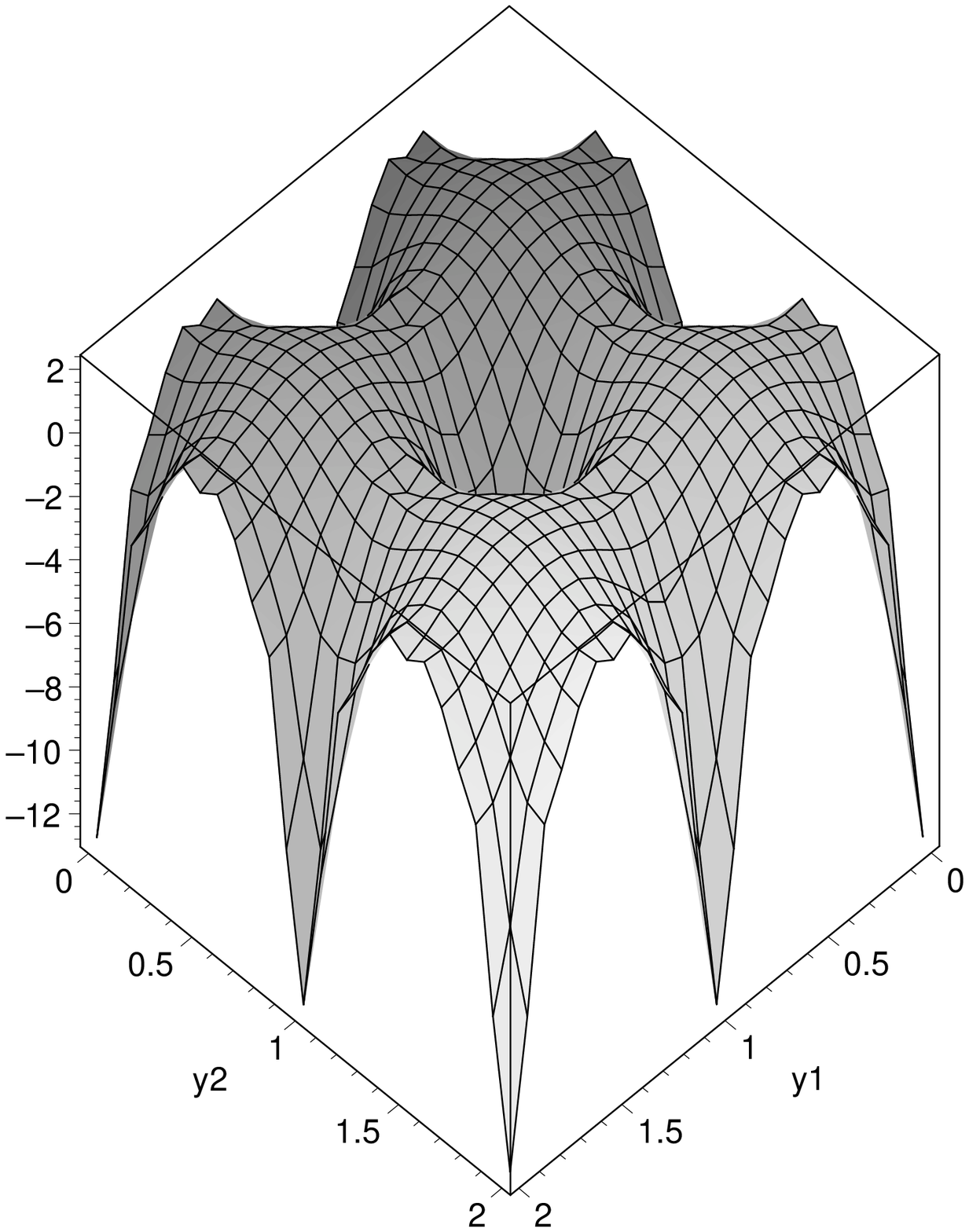,width=6cm}\hspace{2cm}
\epsfig{file=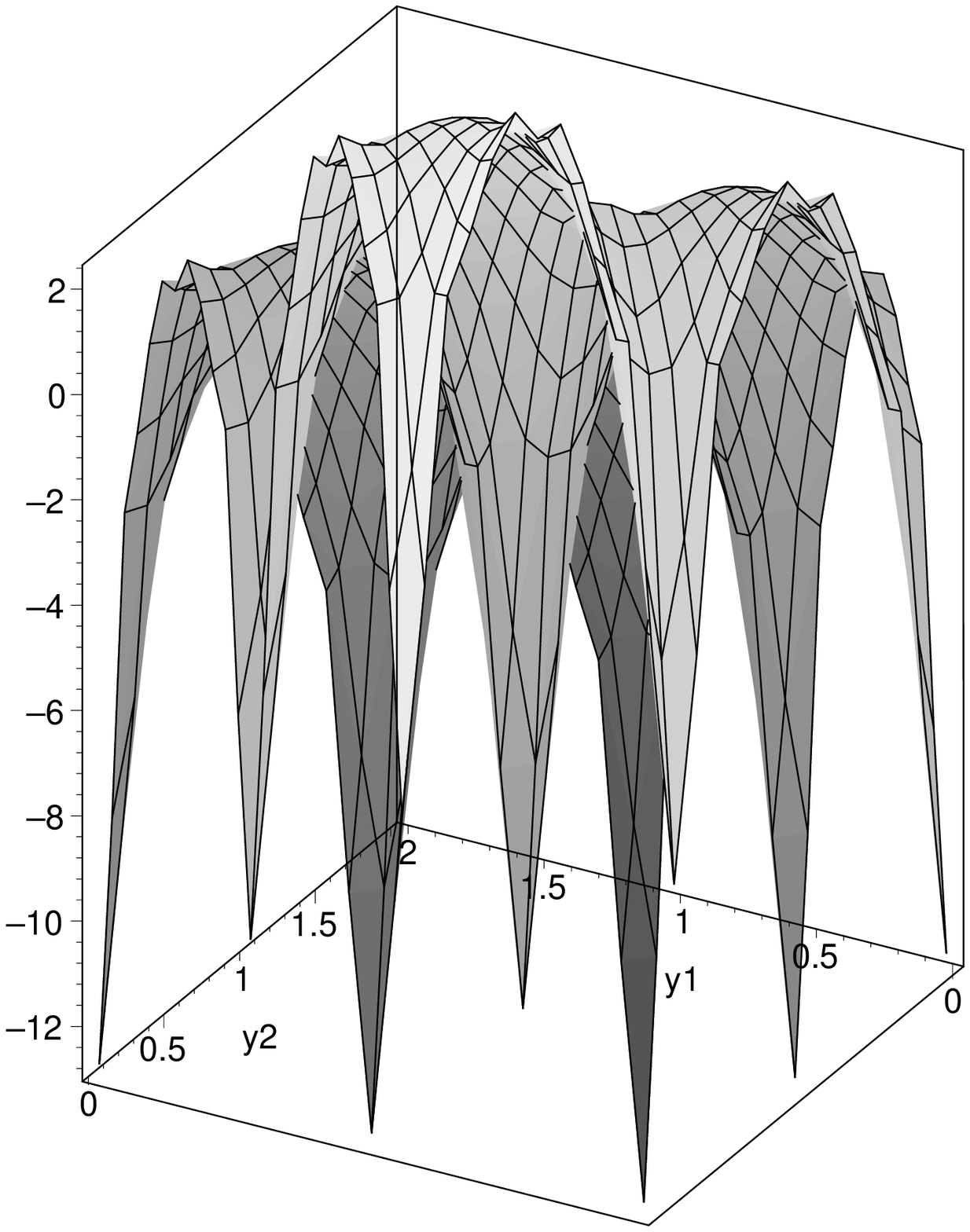,width=6cm} 
\caption{Two views of the effective potential for the inflaton field in
a two dimensional torus transverse to the branes. The periods of the
lattice are normalised to 1 in both directions. Notice how the
rotational symmetry is recovered close to the brane.}
\label{cathedral}
\end{center}\end{figure}

For simplicity, we will consider a particular type of toroidal
compactification, although the details are not very important for our
model. The brane knows about the shape of the compactification space
through the winding modes. When the distance between the D-branes,
$y_i$, is smaller than the compactification radii $R_i$, the sum over
the winding modes can be approximated by:
\begin{equation}
\sum_{m_i} e^{- {t\over2\pi\alpha'}
\sum_i (y_i + m_i 2\pi R_i)^2} \rightarrow e^{- t\,{y^2\over2\pi\alpha'}}
\end{equation}
where $y^2=\sum_i y_i^2$, \ie these winding modes are so massive that
they decouple from the low energy modes or, in other words, it costs a
lot of energy to wind a string around the compact space.\footnote{We
analyse this approximation in greater detail in Appendix C.} To
illustrate this point let us consider a two dimensional compactified
model. The potential is represented in Fig.~\ref{cathedral}. Close
enough to the branes the potential recovers the expected rotational
invariance. The dynamics of the branes close to these points is
described very accurately by the non-compact potential.

\smallskip 

For distances much larger than the string scale $y\gg l_s$, the terms
that contribute most to the integral are those that appear in the
limit $t \rightarrow 0$. In that limit, the partition function
(\ref{ZD4}) becomes $Z(\theta,t) \rightarrow
4\,t^3\,\sin^2\theta/2\,\tan\theta/2$, and the potential, for $l_s \ll
y_i \ll 2\pi R_i$, can be approximated by:
\begin{equation}\label{potstr}
V(y) = 2T_4L -\,{\sin^2\theta/2\,\tan\theta/2\over8\pi^3\alpha'y^2}\,.
\end{equation}
This potential has the expected form, with the right power of the
distance, as could be deduced from Gauss law, i.e. from the exchange of
massless fields in $d_\bot=4$ transverse dimensions. We have plotted
this potential in Fig.~\ref{potential} for the case of $\theta=\pi/12$.

The brane-antibrane system of Refs.~\cite{dss01,bmnqrz01} is the
extreme case of two branes with opposite orientations.  The brane and
antibrane attract each other with a force that depends on the
distance~\cite{bs95}:
\begin{equation}
V(y) = 2T_pL - {B\over y^{d_\bot-2}}\,,
\end{equation}
where $B$ is a positive constant of order one in string units, and
$d_\bot=9-p$ is the number of transverse dimensions to the branes. In
the brane-antibrane system, due to the strong force between them, one
has to chose a special location for the branes in order to have enough
number of \efolds. In our case, as we will show, the number of \efolds
is not so sensitive to the location of the branes, as long as the
angle is sufficiently small.

\EPSFIGURE[ht]{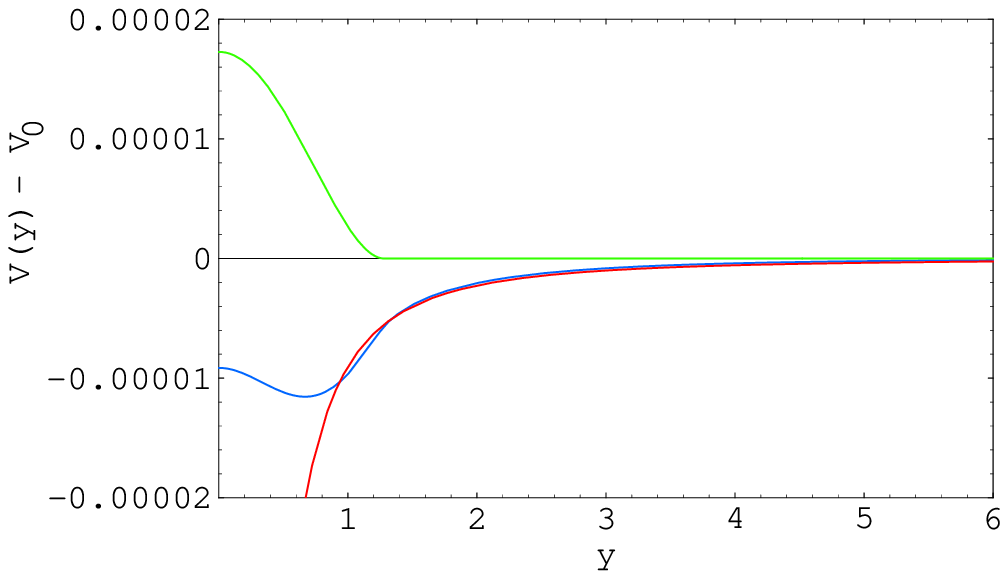,width=9cm}{\label{potential}The attractive
inflaton potential between two D4-branes. The red line corresponds to
the string-derived potential (\ref{potstr}); in blue is the Real part of
the Coleman-Weinberg field-theory limit of the same potential
(\ref{CWpot}), and in green is the Imaginary part, which is non-zero
only when the tachyon condenses, at $y^2 < 2\pi\alpha'\theta$. We have
chosen here $\theta=\pi/12$. The potential $V(y)$ and the distance $y$
are both in units of $\alpha'=1$.}


An alternative, but equivalent, way to obtain the inflaton effective
action at large distances is to start in the closed string picture
from the beginning and consider a probe brane moving in the
supergravity background created by another BPS p-brane solution. The
details of the computation for general Dp-branes are given in the
Appendix A.

The resulting effective action can be organised in a perturbative
expansion of small brane velocities and small supersymmetry breaking
angle $\theta$. The validity of both perturbative expansions are
related, since a small supersymmetry breaking induces a small brane
velocity. Indeed, from the Appendix B, we can see that the ${\cal
O}(v^4)$ terms are negligible within the supergravity regime
$\alpha'/y^2 \ll 1$. Here we just present the first $\theta$-dependent
corrections to the brane probe effective action:
\begin{eqnarray}
S_{\rm probe} = -(T_4L) \int d^4x\,\sqrt{-g}  
&&\left[ \half\left(1+\frac{g_s \pi\sin\theta}{2(M_sL)(M_sy)^2}\right)
|\partial y|^2 \right. \nonumber \\ 
&& - \left. 2\left(1
-\frac{g_s \pi \tan^3\theta}{8(M_sL)(M_sy)^2}\right)\right]\,,
\end{eqnarray}
whose potential coincides precisely with Eq.~(\ref{potstr}), in the
small angle approximation.

\newpage

\section{The inflationary scenario}

In this section we will describe in some detail the constraints that a
generic model of inflation in the brane, and its subsequent
cosmological evolution, should satisfy in order to agree with
observations. We will then use the brane inflation model described
above to obtain phenomenological constraints on the parameters of the
model.

\subsection{Phenomenological constraints}

We will give here the most relevant contraints that should be satisfied
in any scenario of string brane inflation and its subsequent cosmology.
For reviews see \cite{Lindebook,LythRiotto,JGB}.

\begin{enumerate}
  
\item {\em Inflation should be possible}, \ie the energy density of
the universe should be such that the scale factor accelerates: $\ddot
a/a > 0$, or, in terms of the rate of expansion $H=\dot a/a$, we
should have $\epsilon = - \dot H/H^2 < 1$.
  
\item {\em Sufficient number of \efolds during inflation},
$N=\ln(a_{\rm end}/a_{\rm ini})$, in order to solve the horizon and
flatness problems. This constraint depends on the scale of
inflation. In order for the universe to be essentially flat, with
$\Omega_0=1.0\pm0.1$, we require $(\Omega_0-1)/\Omega_0 = \exp(-2N)
(T_{\rm rh}/T_{\rm eq})^2\,(1+z_{\rm eq}) \simeq 0.1$, which implies
\begin{equation}
N \geq 54 + \ln\Big({T_{\rm rh}\over10^{12}\,{\rm GeV}}\Big)\,.
\end{equation}
For the case of GUT-scale inflation, one needs $N \geq  60$, while 
for EW-scale inflation, $N \geq 34$, where the number of \efolds is
computable in terms of the variations in the scalar field that drives
inflation,
\noindent
\begin{equation}\label{Ne}
N = \int Hdt = {1\over \Mp^2} \int_{\phi_e}^{\phi_N} 
{V(\phi)\,d\phi\over V'(\phi)}\,,
\end{equation}
with $\Mp=(8\pi G)^{-1/2}=2.4\times10^{18}$ GeV the Planck mass.

\item {\em Amplitude and tilt of scalar density perturbations} and the
induced temperature aniso\-tropies of the microwave background.
Quantum fluctuations during inflation leave the horizon and imprint
classical curvature perturbations on the metric, which later (during
radiation and matter eras) enter inside our causal horizon, giving
rise through gravitational collapse to the large scale structure and
the observed temperature anisotropies of the cosmic microwave background
(CMB).

The slow-roll parameters are defined as
\begin{eqnarray}\label{epsilon}
\epsilon &=& {\Mp^2\over2} \Big({V'(\phi)\over V(\phi)}\Big)^2 \ll 1 \\
\eta &=& \Mp^2 {V''(\phi)\over V(\phi)} \ll 1 \label{eta}
\end{eqnarray}
The density contrast ($\delta={\delta\rho\over\rho}$) and the
scalar tilt at horizon satisfy~\cite{COBE,Boomerang}
\begin{eqnarray}\label{deltaH}
\delta_H &=& {2\over5}{\cal P}_{\cal R}^{1/2} = {1\over5\pi\sqrt3} 
{V^{3/2}(\phi)\over V'(\phi)\Mp^3} = 1.91\times10^{-5} \,, 
\hspace{1cm} 90\%\ {\rm c.l.} \\
n - 1 &=& {\partial\ln{\cal P}_{\cal R}(k)\over
\partial\ln k} \simeq 2\eta - 6\epsilon\,, \hspace{1cm} 
|n-1| < 0.10  \,, \hspace{1cm} 90\%\ {\rm c.l.}  \label{tilt}
\end{eqnarray}

\item {\em Amplitude and tilt of gravitational waves}. Not only scalar
  curvature perturbations are produced, but also transverse traceless
  tensor fluctuations (gravitational waves), with amplitude and tilt:
\begin{eqnarray}\label{Pg}
{\cal P}_g^{1/2} &=& {\sqrt2\over\pi}{H\over \Mp}   < 10^{-5} 
\hspace{2cm}  {\rm  CMB\ bound}\,, \\
n_T &=& {\partial\ln{\cal P}_g(k)\over
\partial\ln k} \simeq - 2\epsilon  \ll 1  \hspace{1cm}  
{\rm  (not\ yet\ observed).}
\end{eqnarray}

\item {\em Graceful exit}. One must end inflation and enter the
radiation dominated era. This typically occurs by the end of
slow-roll, like in the usual chaotic inflation
models~\cite{Lindebook}, when $\epsilon_{\rm end} = 1$, and {\em not}
when $\eta_{\rm end} = 1$, as incorrectly stated in the literature.
One should then compute the number of \efolds from this time
backwards, to see if there are sufficient \efolds to solve the horizon
and flatness problems.
  
A different way to end inflation is through its coupling to a
tachyonic (Higgs) field, where the spontaneous symmetry breaking
triggers the abrupt end of inflation, like in hybrid
inflation~\cite{hybrid}. This allows for $\epsilon_{\rm end} \ll 1$.
  
\item {\em Reheating before primordial nucleosynthesis}.  Reheating is
  the most difficult part in model building since we don't know to
  what the inflaton couples to. Eventually one hopes everything will
  thermalise and the hot Big Bang will start. One thing we know for
  sure is that the universe must have reheated before primordial
  nucleosynthesis ($T_{\rm rh} > 1$ MeV), otherwise the light element
  abundances would be in conflict with observations. However, since
  the scale of inflation is not yet determined
  observationally,\footnote{The observed amplitud of temperature
    anisotropies in the CMB only gives a relation between the scale
    and the slope of the inflaton potential.} we are allowed to
  consider reheating the universe just above a few MeV.

\item {\em The matter-antimatter asymmetry of the universe}. The
universe is asymmetric with respect to baryon number, and the decays
of the inflaton field typically conserve Baryon number, so it remains
a mystery how the baryon asymmetry of the universe came about.
According to Sakharov, we need B-, C- and CP-violating interactions
out of equilibrium.  The first three occur in the Electroweak theory,
but we would need to reheat the universe above 100 GeV, which may be
too demanding, unless the fundamental Planck scale (in this case the
string scale) is relatively high.
  
\item {\em The diffuse gamma ray background constraints}. If reheating
occurs by emission of massless states to the bulk as well as into the
brane, one must be sure that the bulk gravitons do not reheat at too
high a temperature, because their energy does not redshift inside the
large compact dimensions (contrary to our (3+1)-dimensional world,
where radiation redshifts with the scale factor like $a^{-4}$), and
they could interact again with our (presently cold) brane world and
inject energy in the form of gamma rays, in conflict with present
bounds from observations of the diffuse gamma ray background~\cite{ADD}.
  
\item {\em Model dependent constraints}. In the case of brane
inflation models with large extra dimensions one may prefer that the
low energy effective field theory remains (3+1)-dimensional (otherwise
the cosmological evolution in the brane has to take into account the
evolution of the extra dimensions). In that case, the Hubble scale
should be much larger than the compactification scale, $H\,R \ll 1$. 
This does not impose any serious constraint, in general. Also, in
order to prevent fundamental couplings from evolving during or after
inflation we require that the moduli fields of the compactified space
be fixed.\footnote{We do not provide however any stabilisation
mechanism.}

\end{enumerate}

\subsection{The model of brane inflation}

\EPSFIGURE[ht]{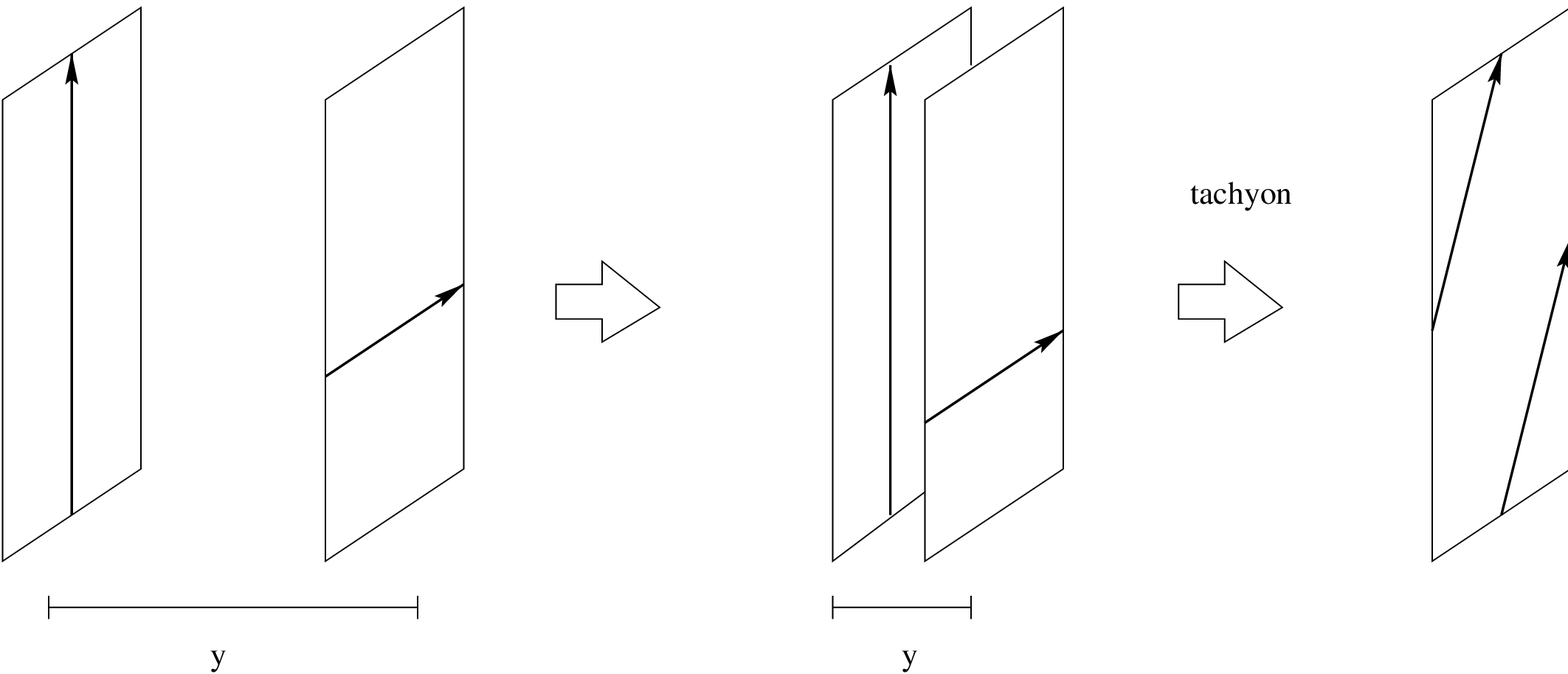,width=10cm}{\label{choque} Two D4-branes
  attract each other to decay in the last step to a bound state. The
  inflation process will take place when the two branes are far away
  if the angle between the two branes is small enough. Note that we
  have chosen here $\theta=\pi/2$ for pictorical purposes. In fact the
  branes are at an angle $\theta\ll1$, so we would expect the final
  brane to be wrapped around the corresponding cycle.}

We will consider here a concrete brane model based on two D4-branes
separated by a distance $y$, and intersecting at a small angle $\theta$,
which attract each other due to the soft supersymmetry breaking induced
by this angle, see Fig.~\ref{choque}. As previously mentioned, we
consider that all the closed string moduli are frozen out. We are also
ignoring the effect of the relative velocity of the branes, which would
introduce a correction to the potential proportional to the fourth power
of the speed. We have computed these corrections in Appendices A and B,
and we have confirmed that they are negligible for our model.  They can
be computed to all order in $\alpha'$ from the one-loop open string
channel. At long distances, only the massless closed string modes
contribute to these corrections, \ie the supergravity approach is
reliable. At short distances, only high speed effects are important, and
they are bigger if the angle between the two branes is small.  But the
process of inflation we are considering occurs at very low velocity and
long distances.  As we will see below, the velocity is related to the
slow-roll parameter $\epsilon$, which in our case is very small.  The
corrections are therefore negligible. These effects are very important
when the distance between the two branes is of the order of the string
scale, then the speed contributions are the dominant ones.

The 4-dimensional effective action can be written as 
\begin{equation}\label{action}
S = \int d^4x\,\sqrt{-g}\Big[\half \Mp^2\,R - 
\half T_4 L (\partial_\mu y)^2 + V(y)\Big]\,,
\end{equation}
where $\Mp^2 = M_s^2(2\pi R\,M_s)^6g_s^{-2}\ $ is the 4-dimensional
Planck mass, and $R$ is related to the compactification volume as $V_6
= (2\pi R)^6$. The potential (\ref{potstr}) for the canonically
normalised inflaton field, $\psi=(T_4L)^{1/2}y$, is given by
\begin{equation}\label{pot}
V(\psi) = M^4\Big(1 - \beta\,{M_s^2\over\psi^2}\Big)\,,
\end{equation}
where $M^4 \equiv 2T_4L = 2(2\pi)^{-4}M_s^4(M_sL)g_s^{-1}$, and the
parameter $\beta$ is a function of the angle $\theta$ between the
branes,
\begin{equation}
\beta = {4\sin^2\theta/2\,\tan\theta/2\over(4\pi)^3}\,,
\end{equation}
which is expected to be very small in order to get a sufficiently flat
(slow-roll) potential.

Let us calculate the derivatives of the potential,
\begin{eqnarray}
\Mp\,{V'(\psi)\over V(\psi)} &\simeq& 
2\beta\,{\Mp\over M_s}\,{M_s^3\over\psi^3} \,,\\
\Mp^2\,{V''(\psi)\over V(\psi)} &\simeq& 
-6\beta\,{\Mp^2\over M_s^2}\,{M_s^4\over\psi^4} \,,
\end{eqnarray}
and the number of \efolds, (\ref{Ne})
\begin{equation}
N = {1\over8\beta}\,{M_s^2\over \Mp^2}\,\Big[{\psi^4\over M_s^4}-
{\psi^4_{\rm end}\over M_s^4}\Big]\simeq{1\over8\beta}\,
{M_s^2\over \Mp^2}\,{\psi^4\over M_s^4}\,,
\end{equation}
in terms of which we can write the slow-roll parameters (\ref{epsilon}) 
and (\ref{eta}), and the scalar tilt (\ref{tilt})
\begin{eqnarray}
\epsilon \simeq {1\over32N^2}\,{\psi^2\over \Mp^2} \,, &&\hspace{1cm}
\eta \simeq -\,{3\over4N} \,,\label{etaN}\\
n &\simeq& 1 -\,{ 3\over2N} = 0.974\,,\label{tiltN}
\end{eqnarray}
which is well within the present bounds from CMB anisotropies, for
$N=54$. The amplitude of scalar metric perturbations (\ref{deltaH})
also gives a constraint on the model parameters,
\begin{equation}
\delta_H = {N^{3/4}2^{-1/4}\over5\sqrt3\pi^3}\,{g_s(M_sL)^{1/2}
\over\beta^{1/4}(2\pi R\,M_s)^{9/2}} = 1.91\times 10^{-5}\,,
\end{equation}
which implies
\begin{equation}
\beta^{1/4} = 3.27\times10^3\,
{g_s(M_sL)^{1/2}\over(2\pi R\,M_s)^{9/2}}\,,
\end{equation}
and thus
\begin{equation}
{\psi_*\over M_s} = (8\beta N)^{1/4}\,(2\pi R\,M_s)^{3/2}\,
g_s^{-1/2} = 1.48\times10^4\,
{(M_sLg_s)^{1/2}\over(2\pi R\,M_s)^3}\,,
\end{equation}
where the asterisc denotes the time when the present horizon-scale
perturbation crossed the Hubble scale during inflation, 54 $e$-folds
before the end on inflation. In terms of the distance between branes,
it becomes
\begin{eqnarray}
{y_*\over l_s} &=& {5.88\times10^5\,g_s\over(2\pi R\,M_s)^3}\,, \\
{y_*\over 2\pi R} &=& {5.88\times10^5\,g_s\over(2\pi R\,M_s)^4}\,, 
\end{eqnarray}
so a compactification radius of order $2\pi R\,M_s = 60$ gives,
for $g_s =0.1$, i.e. in the weak string coupling regime, 
$$l_s < 2.3\ l_s = y_* = {2\pi R\over14} \ll 2\pi R$$
and $\beta^{1/4}
= 7.4\times10^{-6}\,(M_sL)^{1/2}$, which could be made somewhat larger
by chosing a large wrapping length $L$ of the brane around the cycle
in the compactified space, e.g. $LM_s\sim200$, or $\beta^{1/4} \sim
10^{-3}$, in which case the angle for supersymmetry breaking is
$\theta=2\times10^{-3}$. This small angle ensures that inflation will
end, triggered by the tachyon field, when $y \leq y_c = 0.1\ l_s$. This
value of the compactification radius, $2\pi R\,M_s = 60$, gives a
string scale, a Hubble rate and a scale of compactification
\begin{eqnarray}
M_s &=& \Mp\,(2\pi R\,M_s)^{-3}\,g_s \simeq 
9\times10^{12} \ {\rm GeV}\,, \\
M &=& (2M_sL\,g_s^{-1})^{1/4}{M_s\over2\pi} \simeq 
1\times10^{13} \ {\rm GeV}\,, \\
H &=& {M^2\over\sqrt3\Mp} \simeq 
3\times10^7 \ {\rm GeV}\,, \\
R^{-1} &=& 2\times10^{12} \ {\rm GeV}\,.
\end{eqnarray}

Let us now study the production of gravitational
waves with amplitude (\ref{Pg}),
\begin{equation}
{\cal P}_g^{1/2} = {4\over\sqrt3(2\pi)^3}\,
{(M_sLg_s^{-1})^{1/2}\over(2\pi R\,M_s)^6} = 6\times10^{-10}\,,
\end{equation}
which is well below the present bound (\ref{Pg}).

Finally we may ask whether our approximation of using a 4-dimensional
effective theory is correct. For that we need to have the Hubble
radius, $H^{-1}$, of the 4D theory during inflation much larger than
the compactified dimensions,
\begin{equation}
H\,R = {\sqrt2\over(2\pi)^3}\,
{(M_sLg_s^{-1})^{1/2}\over(2\pi R\,M_s)^2} = 1.5\times10^{-5}\,,
\end{equation}
so we are indeed safely within an effective 4D theory.

\subsection{Geometrical interpretation of brane inflation parameters}

\smallskip 
\EPSFIGURE[h]{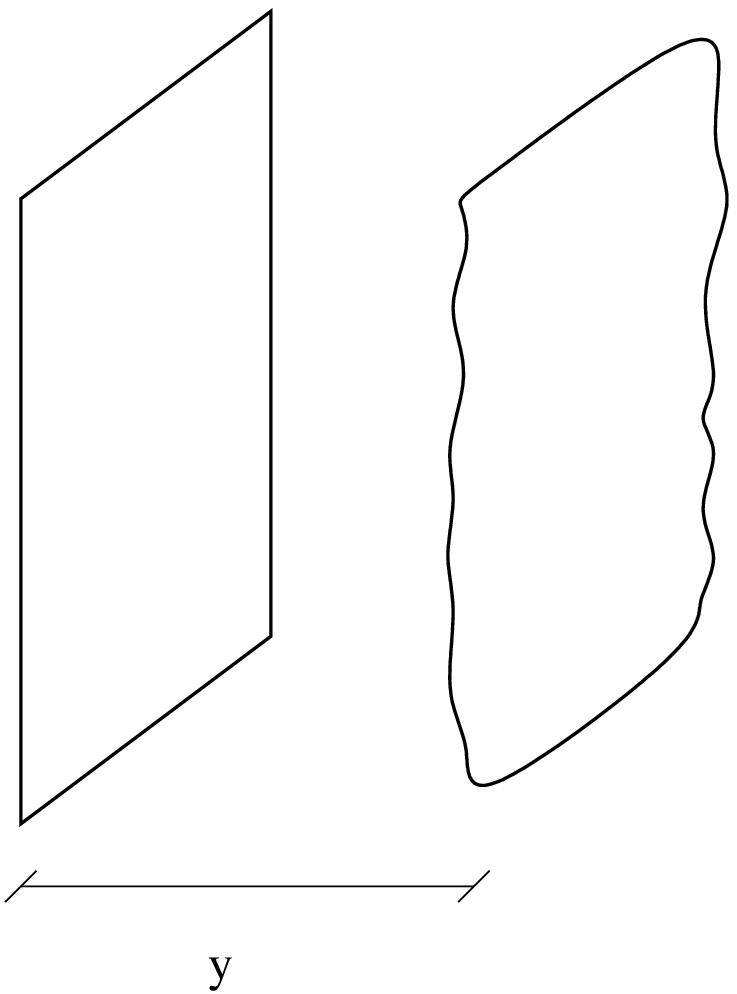,height=4.5cm,width=4.5cm}
{\label{fluctuations} The
dilaton field $y$ is interpreted as the distance between the two branes.
Quantum fluctuations of this field will give rise upon collision to
density perturbations on comoving hypersurfaces. These fluctuations will
be later observed as temperature aniso\-tro\-pies in the microwave
background. \\[5mm] }

Here we will give a \linebreak[3] geometrical interpretation of the
number of \efolds and the slow-roll parameters in our model. The
epsilon parameter (\ref{epsilon}) is in fact the relative squared
velocity ($v=\dot y$) of the branes in the compact dimensions. Since
$\dot\psi^2= M^4v^2/2$ and $3H^2=M^4/\Mp^2$, we have
\begin{equation}
\epsilon=-{\dot H\over H^2}={\dot\psi^2\over2\Mp^2H^2}={3\over4}\,v^2\,.
\end{equation}

The number of \efolds (\ref{Ne}) can be seen to be proportional
as the distance between the branes in the compactified space,
\begin{equation}
N = \int{d\psi\over \Mp\sqrt{2\epsilon}}={H\over v}\,\int dy\,.
\end{equation}
Finally, the eta parameter (\ref{eta}) is the acceleration of the branes
with respect to each other due to an attractive potential of the type
$V(y)\propto y^{2-d_\bot}$, coming from Gauss law in $d_\bot$ transverse
dimensions,
\begin{equation}
\eta = - {d_\bot-1\over d_\bot\,N}\,,
\end{equation}
which only depends on the dimensionality of the compact space
$d_\bot$. Note that the spectral tilt of the scalar perturbations
(\ref{tilt}) therefore depends on both the velocity and acceleration
within the compact space, and is very small in our model of spontaneous
supersymmetry breaking, which makes the branes approach eachother very
slowly, driving inflation and giving rise to a scale invariant
spectrum of fluctuations.

In fact, the induced metric fluctuations in our (3+1)-dimensional
universe can be understood as arising from the fact that, due to quantum
fluctuations in the approaching D4-branes, inflation does not end at the
same time in all points of our 3-dimensional space, see
Fig.~\ref{fluctuations}, and the gauge invariant curvature perturbation
on comoving hypersurfaces, ${\cal R}_k = \delta N_k = H\,\delta y_k/v$,
is non-vanishing, being much later responsible for the observed spectrum
of temperature anisotropies in the microwave
background~\cite{COBE,Boomerang}.

\section{Effective field theory description at short distances}

It is important to make the matching between the superstring theory at
large distances and the supersymmetric quantum field theory in the
brane, whose dynamics will be important for the reheating of the
universe after inflation. For that purpose, note that we can write
the partition function (\ref{ZD4}) in terms of infinite products,
\begin{eqnarray}\label{Ztheta}
Z(\theta,t) &=& {(z-1)^4\over z}\,\Big(\sum_{n=0}^\infty z^{2n}\Big)\,
\prod_{m=1}^\infty {(1-q^m z)^4(1-q^m z^{-1})^4\over(1-q^m)^6
(1-q^m z^2)(1-q^m z^{-2})} \,,\\
q &=& e^{-2\pi t}\,, \hspace{1cm} z = e^{-\theta t}\,.
\end{eqnarray}
The first factor, $(1-z)^4/z = z^{-1}-4+6z-4z^2+z^3$, gives precisely
the lowest lying ${\cal N}=4$ supermultiplet, including the tachyon,
see table~1 and Fig.~2, with the correct multiplicity and
(bosonic/fermionic) sign. Together with the exponential factor,
$\exp(-t\,y^2/2\pi\alpha')$, in Eq.~(\ref{potD4}), it gives the masses
for the one-loop potential
\begin{eqnarray}\label{1loop}
V_{\rm 1-loop} &=& {-1\over(8\pi^2\alpha')^2}\,\int_0^\infty {dt\over t^3}\,
\sum_i\,(-1)^{F_i}\,e^{-2\pi\alpha' t\ m_i^2} \\ 
&=& {1\over64\pi^2}\sum_i (-1)^{F_i}\,m_i^4\,\log\,m_i^2\,, \label{CWpot}
\end{eqnarray}
which corresponds to the Coleman-Weinberg potential for the low energy
effective field theory. The fact that $\sum_i (-1)^{F_i}\,m_i^{2n} = 0,\
(n=1,2,3)$, ensures that (\ref{1loop}) is finite, as it should be, since
supersymmetry is being spontaneously broken.

We can then consider the next series of states in (\ref{Ztheta}). The
factor $\sum_{n=0}^\infty z^{2n} = 1 + z^2 + z^4 + z^6 + \dots$
corresponds to the Landau levels induced, in the dual picture, by the
supersymmetry breaking flux associated to the angle $\theta$. They give,
at any order $N$, a supermultiplet with $\sum_i (-1)^F\,m_i^{2n} = 0,\
(n=1,2,3)$, so they still provide a finite one-loop potential
(\ref{CWpot}).  For a given supersymmetry-breaking angle $\theta$, one should
include in the low energy effective theory the whole tower of Landau
levels up to $N=1/\theta$. For instance, the spectrum for $\theta=\pi/2$
is derived from $(1-z)^4/z\,(1 + z^2 + z^4)=
z^{-1}-4+7z-8z^2+8z^3-8z^4+7z^5-4z^6+z^7$, with masses given by
Table~\ref{mass2}.

\TABLE[ht]{
\begin{tabular}{|c|c|} \hline
Field & $2\pi\alpha'm^2 - \frac{y^2}{2\pi\alpha'}$ \\
\hline
\hline
1 scalar &  $7\theta$ \\
\hline 
2 massive fermions &   $6\theta$ \\
\hline 
3 scalars &   $5\theta$ \\
\hline 
2 massive gauge fields &   $5\theta$ \\
\hline 
4 massive fermions  &   $4\theta$ \\
\hline 
4 scalars &   $3\theta$ \\
\hline 
2 massive gauge fields &   $3\theta$ \\
\hline 
4 massive fermions &   $2\theta$ \\
\hline 
3 scalars  &   $\theta$ \\
\hline 
2 massive gauge fields &   $\theta$ \\
\hline 
2 massive fermions (massless for $y=0$) &   $0$ \\
\hline 
1 scalar (tachyonic for $y=0$) &   $- \theta$ \\
\hline 
\end{tabular}\label{mass2}
\caption{The mass spectrum of the lowest 3 Landau levels, which fall 
into ${\cal N} = 4$ supermultiplets.}
}

Finally, one could include also the first low-lying string states,
whose masses are determined by the expansion of the infinite products
in (\ref{Ztheta}),
\begin{equation}
Z(\theta,t) = {(z-1)^4\over z}\,\Big(\sum_{n=0}^\infty z^{2n}\Big)\,
\Big[1 + {(1-z)^4\over z^2}\,q^2 + {(1-z)^4(1+7z^2+z^4)\over z^4}\,q^4
+ \dots \Big]\,.
\end{equation}
Their structure still comes in ${\cal N}=4$ supermultiplets, so they
again give a finite Coleman-Weinberg potential (\ref{CWpot}).

We will use the whole tower of Landau levels in the effective field
theory to connect the one-loop potential at short distances, determined
by the Coleman-Weinberg potential (\ref{CWpot}), with the full string
theory one-loop potential coming from exchanges of the massless string
modes at large distances, responsible for inflation. This connection
will be essential for the latter stage of preheating and reheating,
because it will provide the low energy effective masses, and the
couplings between the inflaton field $y$ and the effective fields
living on the D4-brane.

The potential (\ref{1loop}) is finite if there are no massless or
tachyonic fields.  When the tachyon appears, \ie at distances smaller
tham $y_c$, there is an exponentially divergent amplitude for $m_i<0$.
A possible strategy to attach physical meaning to this divergence is
to analytically continue the potential (\ref{potD4}) in the complex
$y$-plane.  After the continuation, there is a logarithmic branch
point at $y=y_c$. In this way, we get rid of the divergence and the
potential develops a non-vanishing imaginary part for $y<y_c$, which
signals the instability of the vacuum.

We have plotted in Fig.~\ref{potential} the attractive potential $V(y)$
between two D4-branes at an angle $\theta=\pi/12$. The large distance
behaviour $y^2 \gg \alpha'$ is determined from the supergravity
amplitude (\ref{potstr}), in red, while the short distance potential is
obtained from the Coleman-Weinberg potential corresponding to the
lowest-lying effective fields, which fall into ${\cal N} = 4$
supermultiplets. The real part of the Coleman-Weinberg potential is
drawn in blue in Fig.~\ref{potential}, while the imaginary part is in
green.

In the field theory limit ($q\to0$ and $z\neq0$) we can consider the
infinite tower of Landau levels, while ignoring the string levels, \ie
$Z(\theta,t) \simeq (z-1)^4\,\sum_{n=0}^\infty z^{2n-1}$.  The
finite low energy effective potential up to Landau level $N$ can be
written as
\begin{eqnarray}
V_N(y,\theta)\!&=& {1\over32\pi^2}\,\Big[(y^2-\theta)^2\ln(y^2-\theta)-
4y^4\ln\,y^2+7(y^2+\theta)^2\ln(y^2\!+\theta) \nonumber \\
&-& 8\sum_{n=2}^{2N}(-1)^n(y^2+n\theta)^2\ln(y^2+n\theta)  
+7(y^2\!+\!(2N+1)\theta)^2\ln(y^2\!+\!(2N+1)\theta) \label{VN} \\
&-& 4(y^2\!+\!(2N+2)\theta)^2\ln(y^2\!+\!(2N+2)\theta)+ 
(y^2\!+\!(2N+3)\theta)^2\ln(y^2\!+\!(2N+3)\theta)\Big] \nonumber
\end{eqnarray}
where $y$ stands for $y/(2\pi\alpha')$ and $\theta$ for $\theta/(2\pi\alpha')$.

From this expression we can compute what is the value of the potential
at the distance $y=0$; we expect the tachyon to give an imaginary
contribution to the vacuum energy, which could be interpreted as the
rate of decay of the false vacuum towards the minimum of the tachyon
potential. On the other hand, the real part can be summed over,
\begin{eqnarray}
\label{series}
V_N(0,\theta) &=& {1\over32\pi^2}\,\Big[i\pi 
- 8\sum_{n=2}^{2N}(-1)^n n^2\ln(n)  
+7(2N+1)^2\ln(2N+1) \label{VN0} \\
&-& 4(2N+2)^2\ln(2N+2)+ (2N+3)^2\ln(2N+3)\Big]
\left({\theta\over2\pi\alpha'}\right)^2     \,.         \nonumber
\end{eqnarray}
The infinite sum converges in the limit $N\to\infty$ to:
\begin{equation}\label{V0infty}
V_\infty(0,\theta) = {i\pi-1.70638\over32\pi^2}\,
\left({\theta\over2\pi\alpha'}\right)^2\,.
\end{equation}
This quantity corresponds to the difference between the false vacuum
energy at large distances between the branes, $E_0=2T_pL$, and the
height of the tachyon potential at zero distance ($y=0$), and should
be compared with $\Delta V_{\rm tachyon}$ in Eq.~(\ref{V0}), the
difference between $E_0$ and the final energy density in the tachyonic
vacuum $E_f$. Both are proportional to $\theta^2$, as expected, since
at $y=0$ the one-loop potential (\ref{1loop}) in the field theory limit
$q\to0$ can be written as
\begin{equation}
{-1\over(8\pi^2\alpha')^2}\,\int_0^\infty {dt\over t^3}\,
{(1-z)^4\over z}\,\sum_{n=0}^\infty z^{2n}=
- \Big({\theta\over8\pi^2\alpha'}\Big)^2\,\int_0^\infty {du\over u^3}\,
e^u\,(1-e^{-u})^4\,\sum_{n=0}^\infty e^{-2nu}\,.
\end{equation}
What we have done to obtain (\ref{V0infty}) is to regularise this
integral, for instance introducing a mass cut-off, to make it absolutely
convergent and then commute it with the sum.  The fact that the one-loop
effective action is finite for each Landau level allows us to send the
cut-off to infinity such that only the series (\ref{series}) remains.

\smallskip 
\EPSFIGURE[h]{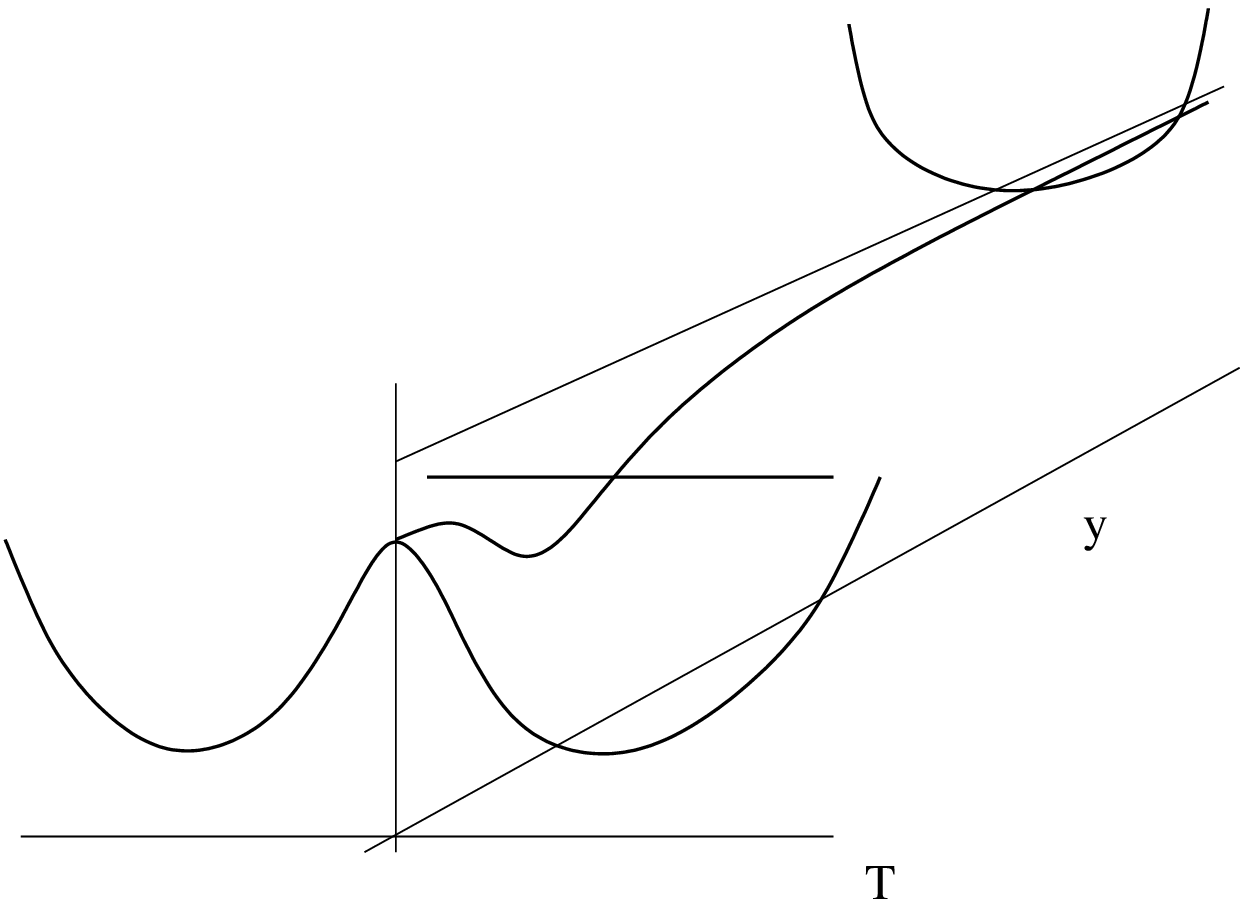,height=5cm,width=5.2cm}
{\label{inflatontachyon} A sketch of the inflaton-tachyon potential
$V(T,y)$. The flat region corresponds to the inflationary regime. }

We have to compare the two quantities: 
\begin{eqnarray}
\Delta V_\infty &=& {1.70638\over32\pi^2}\,
\left({\theta\over2\pi\alpha'}\right)^2\,, \\
\Delta V_T &=& {2M_sLg_s^{-1}\over32\pi^2}\,
\left({\theta\over2\pi\alpha'}\right)^2\,, 
\end{eqnarray}
in the small angle approximation. In order to ensure that the tachyon
minimum is a global minimum for the low-energy effective theory, we need
$M_sL > 0.8352\,g_s$, which is easy to accommodate within the model. A
sketch of the potential in both inflaton and tachyon directions is shown
in Fig.~\ref{inflatontachyon}. The dotted line indicates where inflation
ends and the tachyon instability sets in.

Note that after both branes collide, a vacuum energy density remains,
which, for small angles, is of the same order as the original $V_0$.
The cancellation of this energy density depends on the expectator
branes/orientifolds necessary to cancel the R-R tadpoles and is
irrelevant for the period of inflation, but should be taken into
account at reheating.

\section{Preheating and reheating}

In this section we will briefly describe reheating after inflation,
\ie the mechanism by which the inflaton potential energy density gets
converted into a thermal bath at a given temperature. The details of
reheating lie somewhat out of the scope of the present paper. We will
give here a succint account of what should be expected and leave for the
next paper a more detailed description.

Inflation ends like in hybrid inflation, still in the slow-roll regime,
when the string-tachyon becomes massless and the tachyon symmetry is
broken immediately after. In Ref.~\cite{SSB} it was shown that this
typically occurs very fast, within a time scale of order the inverse
curvature of the tachyon potential, $t_* \sim m_T^{-1} \equiv (\theta/
2\pi\alpha')^{-1/2}$. From the point of view of the low energy effective
field theory description, this is seen as the decay rate (per unit time
and unit volume) or the imaginary part of the one-loop energy density
(\ref{V0infty}).

Assuming that the false vacuum energy $E_0\simeq2T_pL=M^4$ of the two
branes is all of it eventually converted into radiation, we can
compute the reheating temperature of the universe as
\begin{equation}
T_{\rm rh} \simeq \Big({30\over\pi^2g_*}\Big)^{1/4}\,M =
2.2\times10^{12} \ {\rm GeV}\,,
\end{equation}
where we have taken $g_* \sim 10^3$ for the number of relativistic
degrees of freedom at reheating, and we have neglected the energy lost
in the expansion of the universe from the end of inflation to the
time of reheating, since the rate of expansion at the end of
inflation $H \sim M^2/\sqrt3\Mp$ is negligible compared with $m_T$.

The actual process of reheating is probably very complicated and there
is always the possibility that some fields may have their occupation
numbers increased exponentially due to parametric resonance~\cite{KLS}
or tachyonic preheating~\cite{SSB}. Moreover, a significant fraction
of the initial potential energy may be released in the form of
gravitational waves, which will go both to the bulk and into the
brane.  Fortunately, since the fundamental gravitational scale, $\Mp$,
in this model is large enough compared with all the other scales, the
coupling of those bulk graviton modes to the brane is
suppressed,\footnote{The bulk graviton coupling is suppressed by
  powers of $\Mp$, not $M_s$, which is good, since $\Mp/M_s \sim 10^5$
  in our model.}  so we do not expect any danger with the diffuse
gamma ray background~\cite{ADD}, but a detailed study remains to be
done.

\section{Conclusions}

In this paper, we have analysed the realisation of inflationary models
arising from the dynamics of D-branes departing only slightly from a
supersymmetric configuration. The inflaton field is realised in these
models as the inter-brane distance within a compact space transverse
to the branes. As a first example, we obtained the effective
interaction potential for the inflaton field in the case of two almost
parallel D4-branes.  Due to the small supersymmetry breaking, the
potential is almost flat, and therefore satisfies the slow-roll
conditions. It is also interesting to point out the geometrical
interpretation for various other cosmological parameters, such as the
number of \efolds, or the slow-roll parameters $\epsilon$ and $\eta$,
as well as the quantum fluctuations that give rise to CMB anisotropies.

We have analysed the period of inflation in detail within the
supergravity regime. Both D-branes attract each other with a small
velocity at distances much larger than the string scale, but still much
smaller than the compactification scale. In this way, the process
involved is essentially local, without much dependence on the type of
compactification. We find that a sufficient number of \efolds to solve
the flatness and horizon problems can easily be accommodated within the
model. For a concrete compactification radius in units of the string
scale, $2\pi R=60\ l_s$, we find a mass scale $M \sim M_s \sim 10^{13}$
GeV. Moreover, the radius of compactification turns out to be $R^{-1}
\sim 10^{12}$ GeV. An account of the reheating mechanism after inflation
remains to be studied in detail, and in particular the ratio of energy
which is radiated to the bulk versus that into the brane.

A nice feature of the model is the fact that the tachyonic instability
inherent to the brane model triggers the end of inflation.  Since the
D4-branes wrap different non-trivial homology cycles, a single
supersymmetric D4-brane remains after both branes collide. We analysed
the brane interaction at short distances by taking the effective field
theory description on the brane.  We computed the Coleman-Weinberg
potential and obtained a good matching with the supergravity potential.
It remains to study the reheating involved in this regime. We believe
that since the supersymmetry breaking on the brane field theory is
small, maybe it will be possible to study the interactions of the
tachyon with other fields by standard supersymmetric field theory
methods.

There are a series of important issues which should be analysed
\cite{brz02}: i) The first issue is the cancellation of the R-R
tadpoles. Since the branes move in a compact space, the total R-R charge
should vanish.  We can put another D4-brane, wrapping the final cycle
with opposite orientation, or an orientifold plane with exactly the same
opposite R-R charge, far away from the two branes driving inflation,
such that during the period of inflation and the reheating of the
universe, this extra brane is an expectator where the R-R flux can end.
ii) The second issue is the moduli stabilisation. There is a
non-supersymmetric back-reaction on the bulk, as well as non-zero NS-NS
tadpoles, that produces a non-trivial temporal evolution for the closed
string moduli, such us the dilaton and the compactification radii. The
idea is that the time scale involved for this process is much larger
than the time involved during inflation and reheating.  One can compute
the tadpoles by taking the square root of the annulus amplitude. In
general terms, the tadpoles are proportional to
$\sqrt{V_\parallel/V_\perp}$, where $V_{\parallel}$ and $V_{\perp}$ are
the volumes of the compact spaces parallel and perpendicular to the
brane, respectively.  Therefore, for large enough compactification
radii, the tadpoles can be neglected. This is exactly our original
configuration at the beginning of inflation, so we hope to being able to
freeze the moduli, at least during the inflationary period.

Finally, note that from the general expression of the tadpole
potential we can expect that the universe will naturally evolve
towards decreasing both the string coupling and the parallel volume
(to minimise the brane energy), while increasing the transverse
volume. However, in order to precisely quantify this possibility, one
would have to solve the cosmological equations of motion for this
higher dimensional model. We will leave this discussion for our
forthcoming publication \cite{brz02}.

\newpage

\appendix

\section{Probe brane effective action}

Following the conventions of Ref.~\cite{p98}, the type II supergravity
solution for an extremal Dp-brane extended in the directions $x^{\hat\mu}$,
for $\hat{\mu}=0,1,...,p$, is
\begin{eqnarray}
ds^2_{\rm string} && = H^{-\half}_p(x_\bot) ds^2_{p+1}
+ H_p^\half(x_\bot)dx_\bot^2 \,,
\label{Metric}
\\
 e^{\phi} && = H_p^{\frac{3-p}{4}}(x_\bot)\,,
\label{dilaton}
\\
 A_{01...p} && = 1-H_p^{-1}(x_\bot)\,,
\label{RRpotential}
\end{eqnarray}
with all the other supergravity fields vanishing. 
The function $H_p$ is harmonic in the $9-p$ transverse coordinates
$x^I$, $I= p+1,...,9$, with the boundary condition $H_p \to 1$ for
$|x_\bot| = \sqrt{x^I x_I} \to \infty$, in order to recover the flat Minkowski
spacetime for the metric (\ref{Metric}). Considering the rotational invariance
in the transverse space, we have
\begin{eqnarray}
H_p(|x_\bot|) && = 1 + \frac{c_p g_s N}{(M_s |x_\bot|)^{7-p}}\,,
\label{harmonicH}
\\
c_p && = (2 \sqrt{\pi})^{5-p} \,\Gamma (\frac{7-p}{2})\,.
\end{eqnarray}
Our case corresponds to $p=4$ and $N=1$ (number of background branes),
but to stress the generality of this technique, we shall keep $p <7$ and
$N$ arbitrary.

The effective action for a probe Dp-brane moving in this background 
is given by
\begin{eqnarray}
S_{\rm p-brane} &&= S_{\rm BI} + S_{\rm WZ}\,,
\label{probe}
\\
S_{\rm BI} &&=-T_p \int_{\rm Dp} d^{p+1}\xi\ e^{-\phi}
\sqrt{-{\rm det}(G_{MN}(X)\partial_\mu X^M \partial_\nu X^N)}\,,
\\
S_{WZ} &&= -T_p \int_{\rm Dp} A_5\,.
\end{eqnarray}

We take the configuration for the probe rotated 
in the $\{x^p,x^{p+1}\}$ plane:
\begin{eqnarray}
&&X^\mu = \xi^\mu\,,  \hspace{2cm} \mu = 0,1,...,p-1\,,
\nonumber
\\
&&X^p = \xi^p \cos\theta\,,
\nonumber
\\
&&X^{p+1} =\xi^p \sin\theta\,,
\nonumber
\\
&&X^I =y^I(\xi^\mu)\,, \hspace{1.5cm} I=p+2,...,9 \,.
\label{probe_configuration}
\end{eqnarray}
Plugging into (\ref{probe}), we obtain 
\begin{eqnarray}
S_{\rm BI} = && -T_p \cos\theta \int d^px\,d\xi^p\sqrt{-g}\ 
H_p^{-1}\sqrt{1+H_p \tan^2\theta} 
\sqrt{{\rm det}(\delta^\mu_\nu + H_p \partial^\mu y^I\partial_\nu y_I)}\,,
\label{BIone}
\\
S_{\rm WZ} =&& -T_p \cos\theta \int d^px\,d\xi^p\,\sqrt{-g}\ (1-H_p^{-1})\,,
\label{WZone}
\end{eqnarray}

There are two conditions in order to have that the brane probe action
(\ref{probe}) is a valid approximation for the effective action of the
inflaton field. The first is that the supersymmetry breaking mass scale
is small with respect the string and Plank scales, to aboid relevant
supersymmetry-breaking back-reaction effects on the background.  In our
model, supersymmetry is spontaneously broken by the rotated brane
configuration (\ref{probe_configuration}).  For sufficiently small
angle, we can take the non-supersymmetric back-reaction as a sub-leading
effect, at least in the region of small curvature. 

This brings us to
the second condition, which is to use the action (\ref{probe}) only in
the valid supergravity regime.  It corresponds to large inter-brane
distances with respect to the string length, \ie $M_s|x_\bot|
\gg 1$, in order to neglect the massive stringy effects.  Later on, we
will see that small angle allows us to treat the inflationary period
within the supergravity regime, such that both conditions are satisfied.
For the analysis of preheating, when $M_s|x_\bot| \sim 1$ and the
tachyon is turned on, massive closed string states become relevant and
one should follow a different strategy to describe the relevant degrees
of freedom. We leave the quantitative analysis of preheating for the
future \cite{brz02}.

Notice that the two above mentioned conditions, $\theta/\pi$ and
$(M_s|x_\bot|)^{-1}$ much smaller than one, correspond to the two
natural expansion parameters for the probe action (\ref{probe}).  We
can perform an expansion for the first square-root factor in
(\ref{BIone}) of the form
\begin{equation}
\cos\theta \sqrt{1+H_p \tan^2\theta} = \sqrt{1+I_p} 
= 1 +\frac{I_p}{2} -\frac{I_p^2}{8} + {\cal O}(I_p^3)\,,
\end{equation}
where we have defined
\begin{equation}
I_p = \sin^2\theta \frac{c_p g_s N}{(M_s|x_\bot|)^{7-p}} \ll 1\,.
\end{equation}

For the second square-root in (\ref{BIone}), we can expand with
respect small velocities, or what is the same, in the gradient
$K^{\mu}_{\nu} = \partial^{\mu} y^I \partial_{\nu} y_I$, of the form
\begin{equation}
\sqrt{ 1 + H_p K} = 1 + H_p{\rm tr}K + 
\frac{H_p^2}{4} \left(\half ({\rm tr K)}^2 
- {\rm tr}(K^2)\right) + {\cal O}(K^3) \,.
\end{equation}
The result is that we can organise the kinetic and potential terms for 
the inflaton in a perturbative expansion 
in $\theta/\pi$ and inverse powers of $(M_s|x_\bot|)$:
\begin{equation}
S_{\rm p-brane} = \int_{\rm R^{1,p-1}}d^px\,\sqrt{-g}\ 
({\cal L}_{K}(y) - V(y))\,,
\label{probe2}
\end{equation}
with the kinetic terms given by
\begin{eqnarray}
{\cal L}_{K}(y) = -T_p &&\int_0^L d\xi^p\ \left(1+\half I_p(y,\xi^p) 
-\frac{1}{8}I_p^2(y,\xi^p) + ...\right)
\nonumber
\\
\times &&\left(\half{\rm tr}K(y) + 
\frac{H_p(y,\xi^p)}{4} \left(\half({\rm tr}K(y))^2 
- {\rm tr}(K^2(y))\right) + ...\right)\,,
\label{probe_kinetic}
\end{eqnarray}
and the potential terms given by
\begin{equation}
V(y) = T_p \cos\theta\int_0^L d\xi^p 
\left(1 +\half \tan^2\theta 
-\frac{1}{8}H_p(y,\xi^p)\tan^4\theta + ...\right)\,.
\label{probe_potential}
\end{equation}
Since the brane probe is rotated with respect the brane background,
the harmonic function $H_p$ (and consequently $I_p$ also) depends on
the world-volume coordinate $\xi^p$ through the combination $|x_\bot|
= \sqrt{y^2 + \xi_p^2 \sin^2\theta}$.  This longitudinal Dp-brane
direction is wrapping a non-trivial cycle in the toric plane
$\{x^p,x^{p+1}\}$.  If we consider a squared torus with size
$4\pi^2 R_\bot^2$, then the lenght of the cycle is $L = 2\pi
R_\bot/\sin\theta$.  Notice that we can also include the corrections
due to a finite compactification scale in the remaining transverse
coordinates, $X^I$ for $I=p+2,...,9$, by just replacing the expression
of the harmonic function $H_p$ in (\ref{harmonicH}) by its
corresponding Kaluza-Klein expression. We can look at the Appendix C
for a discussion on this.  Again, they will be small if we can keep
our analysis for an inter-brane distance smaller than the
compactification scale.

As an illustrative example, let us compute the first
non-supersymmetric corrections to the effective action of the
inflaton.  Form (\ref{probe_kinetic}), the first correction to the
kinetic energy comes from the integral
\begin{eqnarray}
\half \int_0^\infty d\xi^p\ I_p(y,\xi^p) &&=
\frac{\sin\theta c_p g_sN}{2 M_s} \int_0^\infty dx 
\left( M_s^2 y^2 + x^2 \right)^{\frac{p-7}{2}}
\nonumber
\\
&&= \frac{c_{p+1} g_sN}{(M_s y)^{6-p}}\frac{\pi \sin\theta}{2M_s}\,.
\end{eqnarray}
In this case we have for the kinetic term
\begin{equation}
{\cal L}_K = -\frac{T_pL}{2}\left( 1 + \frac{\pi\sin\theta}{2M_s L} 
\frac{c_{p+1}g_s N}{(M_s y)^{6-p}} +...\right)
|\partial y|^2\,,
\end{equation}
and for the potential energy, 
\begin{equation}
V(y) = T_p L \left(1 -\frac{\pi\tan^3\theta}{8 M_sL}
\frac{c_{p+1}g_sN}{(M_sy)^{6-p}} + ...\right)\,,
\end{equation}
For $p=4$, $N=1$ and small angles $\theta\ll\pi$, it gives exactly
half of Eq.~(\ref{potstr}). This should be expected, since we have
only considered one probe brane, and we have not introduced, for
instance, the rest energy of the other.

\section{Velocity-dependent corrections to the inflaton 
potential.}\label{velocitydep}

The corrections due to the speed of the branes can be obtained very
easily from the one-loop open string amplitude (or equivalently by the
tree level exchange of closed strings) by considering two angles. The
first angle is our $\theta$, the second one can be taken to be imaginary
and corresponds to the hyperbolic tangent of the speed. This procedure
is described in detail in \cite{p98}.  The potential at long distances
has a dependence with the velocity of the form:
\begin{equation}
V(y,v) = {-1\over16\pi^3\alpha'r^2} 
\frac{(\cos{\theta} - \gamma)^2}{\gamma \sin{\theta}}  \,,
\end{equation}
where $r^2 = y^2 + \tau^2 v^2$ is the distance between the two branes
and $\gamma$ is the relativistic factor: $\gamma = 1/\sqrt{1-v^2}$.
The same expresion can be obtained from the probe brane effective
action (\ref{probe2}) at leading order in $\alpha'/y^2$. Notice also
that this potential is invariant $v \rightarrow -v$, as expected from
time reversal.

One can expand the above expression in the low speed limit to get the
corrections to the static potential from the motion of the branes:
\begin{eqnarray}
\frac{(\cos{\theta} - \gamma)^2}{\gamma \sin{\theta}} & 
= & \frac{(\cos{\theta} - 1)^2}{\sin{\theta}} 
+ \frac{v^2}{2} \sin\theta - \frac{v^4}{8}
\frac{(\cos^2\theta-3)}{\sin\theta} + {\cal O}(v^6) \,.
\end{eqnarray}
The first term is just the static interaction:
\begin{equation}
V(y,0) = {-1\over16 \pi^3 \alpha' r^2} 
\frac{(\cos{\theta} - 1)^2}{\sin{\theta}} =  
\frac{-1}{8\pi^3 \alpha' r^2} \sin^2{\theta\over2}\,\tan{\theta\over2}\,.
\end{equation}
The second term is the correction to the inflaton kinetic term:
\begin{equation}
V_2(y,v) = \frac{-1}{32 \pi^3 \alpha' r^2} \sin\theta\,.
\end{equation}
This correction vanishes in the supersymmetric case 
when the two branes become parallel \cite{p98}.
The fourth order correction is:
\begin{equation}
V_4(y,v) = \frac{-1}{128 \pi^3 \alpha' r^2} 
\frac{(3 - \cos^2\theta)}{\sin\theta}\,.
\end{equation}
Since our description of the inflationary process occurs at very low
velocities we can estimate the order of magnitude of the correction
due to these terms. It can be easily estimated by considering the
highest speed $v$ associated with the slow-roll parameter $\epsilon$
during the inflationary process. One may worry that at low angles the
sine factor in the denominator could give big corrections.
Fortunately, when substituting the speed $v$ into the above formulae,
one gets negligible corrections to our results.

For completeness, we can take the ultra-relativistic limit $v \rightarrow 1$. 
If we denote $\delta = 1-v$ and expanding the above potential one gets:
\begin{eqnarray}
\frac{(\cos\theta - \gamma)^2}{\gamma \sin\theta} & = &
{1\over\sin\theta \sqrt{2\delta}}
- 2 \cot\theta + {\cal O}(\sqrt\delta)\,.
\end{eqnarray}
Notice that the interaction becomes stronger at high speeds. For a
discussion about these limits, see Ref.~\cite{p98}.

\section{Discussion on compact potentials.}
\label{compactaprox}

Within this model we are assuming that we are close enough to one of
the D-branes that the effects of the other branes, as well as the
effect of the winding modes around the compact space, are negligible.
This means that the process we are considering is a local one, not
depending very strongly on the details of the compactification and
initial conditions. We will justify here this approximation. 

To evaluate the importance of this approximation, we consider that the
Dirichlet directions are compactified on a torus. In order to make a
more general discussion let us consider $D$ Dirichlet dimensions in a
torus $T^D$. These directions appear in the cylinder partition
function as a sum over winding modes, see Eq.~(\ref{ZD4}). Commuting the
integral with the sum one can express the potential as a sum over the
images of the brane on the torus. Thus one obtains at distances
bigger than the string length an effective potential of the form:
\begin{equation}
V(y_i) = \sum_{n_i} \frac{-k}{\Big[\sum_i(y_i + L_i n_i)^2
\Big]^{\frac{D-2}{2}}} \,,
\end{equation}
where $k$ is a constant that will depend on the angles, fluxes, etc.
When the system is supersymmetric the $k$ vanish, \ie there are
no interactions.

This formula is valid if the number of Dirichlet torroidal dimensions
is different from two. In the case $D = 2$ the integral produces a
logarithmic potential. It is easy to see that the long distance
behaviour is divergent and one should try to regularise the above
expression. This divergence can be undertood very easily. It is due to
the propagation of massless closed strings fields at long distances,
i.e. an IR divergence. From the formulae below one can see that it is
related to the NS-NS uncancelled tadpoles. If these tadpoles are not
present one expects this divergence to disappear. That is what happens
in the supersymmetric case, where the constant $k$ is zero, and in the
non-compact case, where the images are not present.  The computation
can be done very easily by expressing the above sum as an integral:
\begin{equation}
V(y_i) = \frac{-k}{\Gamma(\frac{D-2}{2})} 
\int_0^{\infty} \frac{dt}{t} \, t^{\frac{D-2}{2}} \sum_{n_i} 
e^{-t\,\sum_i(y_i + L_i n_i)^2}\,,
\end{equation}
which can be expressed, through the change of variables $t=-\pi/x$, in
terms of Elliptic functions as
\begin{equation}
V(y_i) = \frac{-k\,\pi^{\frac{D-2}{2}}}{\Gamma(\frac{D-2}{2}) V_T}
\int_0^\infty dx \, \prod_{i=1}^D
\theta_3\Big(\frac{y_i}{L_i},\frac{ix}{L_i^2}\Big)\,,
\end{equation}
where $V_T=\prod L_i$ is the volume of the torus. This expression diverges
quadratically for $x \rightarrow \infty$, and can be regularised by
subtracting a coordinate-independent piece in the integral. That is like
subtracting an infinite constant. The regularised expression becomes
\begin{equation}
V(y_i)_{\rm reg} = \frac{-k \,\pi^{\frac{D-2}{2}}}{\Gamma(\frac{D-2}{2}) V_T} 
\int_0^\infty dx \, \Big[\prod_{i=1}^D \theta_3\Big(\frac{y_i}{L_i},
\frac{ix}{L_i^2}\Big)- 1\Big] \,.
\end{equation}
By commuting again the integral and the sum one can compute the
integral for each of these terms and one gets:
\begin{equation}
V(y_i)_{\rm reg} = \frac{-k\, \pi^{\frac{D-2}{2}}}{\Gamma(\frac{D-2}{2}) V_T} 
\sum_{m_i \neq 0}\ {\prod_i^D\ \cos(2 \pi m_i y_i/L_i) \over 
\sum_i^D\ m_i^2/L_i^2} \,,
\end{equation}
where the term with all the $m_i = 0$ has been subtracted out.
Another way of interpreting this regularisation is by taking the
Fourier transform of the previous one without taking into account the
zero term, i.e. the infinite one. Of course, the constant term does
not affect the dynamics.

\vspace{2cm}



\end{document}